\documentclass{article}

\usepackage[preprint]{nips_2018}

\usepackage[utf8]{inputenc} 
\usepackage[T1]{fontenc}    
\usepackage{hyperref}       
\usepackage{url}            
\usepackage{booktabs}       
\usepackage{amsfonts}       
\usepackage{nicefrac}       
\usepackage{microtype}      
\usepackage{amsmath,amssymb}
\usepackage{graphicx}
\usepackage{subcaption}
\usepackage{epstopdf}
\usepackage{epsfig}

\title{AI based Safety System for Employees of Manufacturing Industries in Developing Countries}

\author{
   A. ~Das, S. ~Panda, S. ~Datta, S. ~Naskar, P. ~Misra, T. ~Chattopadhyay\\
   Tata Consultancy Services Research and Innovation labs, \\
   Kolkata, India \\
\texttt{(abhisek.das2, satanik.panda,suman.datta, soumitra.naskar)@tcs.com} \\
\texttt{and (prateep.misra, t.chattopadhyay)@tcs.com}   
}

\begin{document}

\maketitle

\begin{abstract}
In this paper authors are going to present a Markov Decision Process (MDP) based algorithm in Industrial Internet of Things (IIoT) as a safety compliance layer for human in loop system. Though some industries are moving towards Industry 4.0 and attempting to automate the systems as much as possible by using robots, still human in loop systems are very common in developing countries like India. When ever there is a need for human machine interaction, there is a scope of health hazard. In this work we have developed a system for one such industry using MDP. The proposed algorithm used in this system learned the probability of state transition from experience as well as the system is adaptable to new changes by incorporating the concept of transfer learning. The system was evaluated on the data set obtained from 39 sensors connected to a computer numerically controlled (CNC) machine pushing data every second in a 24x7 scenario. The state changes are typically instructed by a human which subsequently lead to some intentional or unintentional mistakes and errors. The proposed system raises an alarm for the operator to warn which he may or may not overlook depending on his own perception about the present condition of the system. Repeated ignorance of the operator for a particular state transition warning guides the system to retrain the model. We observed 95.61\% alarms raised by the said system are taken care of by the operator. 3.2\% alarms are coming from the changes in the system which in turn used to retrain the model and 1.19\% alarms are false alarms. We could not compute the error coming from the mistake performed by the human operator as there is no ground truth available for that. 

\end{abstract}

\section{Introduction}
Recent advancement in manufacturing industry is highly benefited by the use of computer numerically controlled (CNC) machines which has rejuvenated that industry significantly. High end CNC machines are now in use in most of the developed countries like USA, Germany, japan. But at the same time developing countries like India is not far behind and many such industries in India are using CNC [5]. But the use of CNC should be restricted to a properly trained operator only. The major problem of utilization of such machines in developing countries occurring from lack of such training and thus it leads to a contradiction in safety [8]. In case of such untrained user, the system may undergo some intentional or un-intentional mistakes and errors. Thus there is a need for checking the different sensor values before the machine transits to a state that may interact with the human in system. The need for such safety methods in such a scenario is stated in some of the standards like [2], [3], [4]. The major take away points are (i){\it Protective separation distance:} Shortest permissible distance between any moving hazardous part of the robot system and any human in the collaborative workplace, (ii) {\it Safety design aspects:} Possible stress, fatigue, or lack of concentration, and (iii) {\it Error or misuse:} intentional or unintentional error committed by the operator
The standards also identified that the following tasks may leads to health hazards: (i) {\it Automatic and manual restart of the system}, (ii) {\it Tasks involving more than one operator}, and (iii) {\it Frequency and duration of contact between human and the machine under consideration}

So in the proposed solution we aim to address these issues like checking the distance between the operator and the spindle motor of CNC machine, checking the illegal state transitions. The proposed method also aims to implement a basic transfer learning paradigm by learning automatically in case the alarms raised by the system is repeatedly gets rejected by the operator. On the other hand this adaptability has an impact on the memory and memory crunch may be resulted. Thus the proposed method also used a Least Recently Used (LRU) scheduling to clean the memory of stale data. The contribution of the proposed work is (i) Proposed system checks the sensor values so that the safety of human in loop can be ensured before the state change instruction given by an operator, (ii) Trigram of state transitions are noted initially. Based on these data a probability is assigned to each state transitions, (iii) Incorporate the basic transfer learning concept to make the system adaptable to change. The proposed system has been tested on a stored data set obtained using an IoT data acquisition platform called TCS Connected Universal Platform (TCUP) [7].
\section{Proposed Method}
Proposed method is based on Markov Decision Process (MDP) which is characterized by 5 tuples. $M=(S,A,p,R,\gamma)$ where each of the tuples where (i) S:S defines the set of states, (ii) A: A is the set of actions, (iii) p: p is the transition probability from one state $s_{i}$ to $s_{i+1}$. So this is a function from p->SxS, (iv) R:R is the reward received in the transition of state $s_{i}$ to $s_{i+1}$. This is a function R->SxA. In case some alarm raised by the system is overruled by the human operator is used to update the reward function, and (v) $\gamma$: This is a forgetting factor or discount factor. It is also a function $\gamma$->SxA. This is used to remove the stale data. Now let us describe how each of these tuples are used in the proposed method. In a CNC machine a part is made by following a set of sequence of operations. S is the states of the machine which indicate the different operations performed in a particular sequence. Typically the operations are performed in a sequence but some times it is required to break the normal sequence. For example, before going for a fine cut using the machine it is preferable to check the rough cut and so the rough cut is re-done before the onset of fine cut. So the sequence of operations follow certain patterns. Again each of the operation has a particular pattern in different sensor values and the duration of the operation mostly varies within a small range of values unless there is a major change in the description of the target module. Some of these state transitions require human intervention to provide/remove the material/element for further processing. During these phases, human body parts are getting exposed to some machine parts and may lead to health hazard. So, whenever any state change is instructed, it is verified by the proposed MDP based design. So state transitions are checked using two aspects namely (i) current state of the sensors, (ii) history of allowed state transitions. Action state (A) is used to check the status of the sensor values. For example, Axis-1 Relative Position depicts the distance of the material with the spindle that is rotating at a high speed. So there is a minimum distance requirement from which the operator should operate to avoid contact with the machine. The distance of the material from the spindle is quite high at the beginning and end of operation 7 as the tray comes back to its start potion in case of normal scenarios. But because of error/mistake performed by the operator in certain sequences it is violated mostly as the tray holding the material to cut doesn't come back to its original position at the end of the operation. In the proposed method the system raises alarm in such scenarios. The concept of adaptability to new scenario was described first by Kifer et. al. in [6]. He mentioned that the performance of a system depends on its capability to adapt to changes and new concepts as well as to forget outdated concepts as it may become an overhead for the system. Thus the concept of incremental learning and decremental learning evolved [1]. In our proposed method $R$ is the factor called reward that is used to create an instance of the object when repeated rejection happens. Repeated rejection in a semi supervised method means that the system needs to be trained to the new change in the system. The new instance starts to populate the statistics of the instances where the alarms raised by the system gets rejected by the operator. So, from this time the system starts to check whether the sensor value is marked as the outlier for the newly adapted model as well as for the previously obtained model. But this also creates a overhead for the storage. Thus the time stamp when the model is used is also noted. Finally an least recently used (LRU) algorithm is used to de-allocate the memory for the outdated or stale data models.

\section{Experiments}
In this section we shall initially describe the method of implementation and then will describe the data acquisition method and finally will describe about the sensors. The code has been implemented in Java language $(SDK Version 1.8.0\_171)$ in Ubuntu environment. The code is designed in a modular fashion. Five main modules constitute the code. The modules compute mean, median, mode, standard deviation, minimum and maximum of each sensor data over six months. The code takes around one minute for execution when we give six months data as input.

\begin{figure*}[t]
\includegraphics[width=4 in]{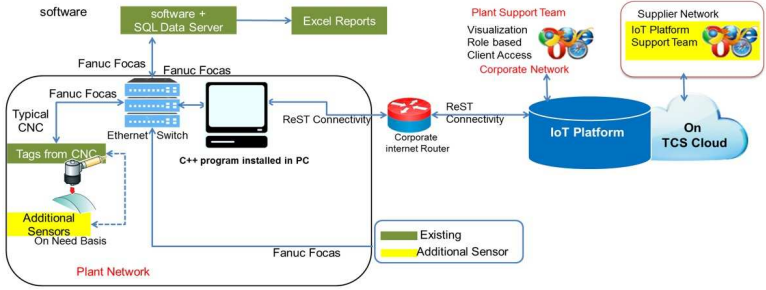}
\caption{Sensor data acquisition from CNC machines}
\label{fig:dataAcquisition}
\end{figure*}

{\it Data Acquisition: } The method we have adapted in this experiment to acquire data from the CNC machines in an IoT platform is depicted in Figure \ref{fig:dataAcquisition}.
The current setup of sensors in the factory follows a logical hierarchy in terms of its location and operation. As part of this project nine CNC machines are taken into consideration. All the quality data captured manually and send to the cloud as a bulk excel file to be stored for analysis. This process produces more bad components repeatedly getting created due to the same error on the process. For the whole month a plan gets created for the production schedule for all the machines and gets shared in the form of an excel which also gets stored on cloud for correlation and analysis. All the jobs are done on those components are metal cutting operations with multiple dimension and directions. For each operation steps on the machine there is a baseline value available from past production data on which the plan is laid out considering the optimum time taken for each steps. Data are transferred via Message Queuing Telemetry Transport (MQTT) protocol for two way communication using REpresentational State Transfer(REST) Application Programming Interface (API)s and JavaScript Object Notation (JSON) payloads. Inside the factory data are generated by this CNC machines multiple sensors. Most of this machines are having Fanuc focus controllers. Data are taken out by directing using Fanuc interface from the micro-controller lever. There is an edge processing layer built which runs a gateway box inside the factory in close proximity of these machine where the data gets extracted from the controller and converted to a MQTT payload then sent to the cloud server interface. The data sending frequency is per second basis for very high changing data elements. For the lower frequency data element data kept in a buffer and accumulated then sent to cloud at one minute frequency. On the cloud the data gets stored on an IoT platform named TCUP. 

{\it Description of the data set: } In the previous section we have described about the method of data acquisition. In this section we are going to describe about the data. The data is stored in a hard drive and all the experiments were performed on it. The data set contains the four axis CNC machine data for the month of March, 2017 to August, 2017 sampled at every second. The data-set includes 52 sensor data out of which 11 may lead to impact on health hazard of the workers are described below. 5 different types of machine generated alarms are also there. All together the data-set contains 12956793 second of data and the size of the data set is 10.8 GB. The sensor data used in the experiments and the physical significance of those sensors are as below: (i) Axis1 Relative Position (Relative  position of the controlled axis 1), (ii) Axis2 Relative Position	(Relative  position of the controlled axis 2), (iii) Axis3 Relative Position	(Relative  position of the controlled axis 3), (iv) Axis4 Relative Position	(Relative  position of the controlled axis 4), (v)Axis1 ServoLoad	(Servo Load of  Axis1), (vi)Axis2 ServoLoad	(Servo Load of  Axis2), (vii)Axis3 ServoLoad	(Servo Load of  Axis3), (viii)Axis4 ServoLoad	(Servo Load of  Axis4), (ix)Feed Data	(Actual feed rate of the controlled axes), (x)Speed Data	(Actual rotational speed of the spindle load meter data), (xi)Converted Spindle Speed	(Converted spindle speed value related to constant surface speed control on CNC). The system generates a confusion matrix with the notation like TN is the scenario where the system identified the data to be faulty but it is overlooked by the operator and the reverse scenario i.e. when the operator thinks the data to be error but no alarm is raised by the system is marked as FN. The ground truth of the system is obtained from the manual operator. 

\subsection{Results}
In this section we are going to describe the experimental results and explain them. The basic statistics of duration for different operations over different months clearly shows that there is certain change in the product description for which the duration of the operations 6, 7, and 10 has changed significantly in the statistics of the month of March and May. This leads to need for retraining of the system. But as the proposed method is adaptable to take care the change in the scenario it takes some time (3 hours) to create a new model. Another interesting fact can be noted is that the standard deviation of duration for each operation is quite high because of mishandling of the operation. For example, we have observed that the machine keeps running even when an operation is completed or sometimes a wrong operation started to execute and needs to be stopped before it is completed. The minimum and the maximum values shows those wrongly handling scenarios. Thus the proposed method raises alarm to stop such misuse by checking the possible state transitions. The data set can not be annotated while the data is getting captured in real time. But some domain experts can mark them as normal and abnormal because of the presence of the alarms obtained from the data. It is also noted that the hardware alarm attached with the system didn't raise any abnormality in more than 90\% scenario which also proves the need for such a system proposed here. From the Table~\ref{tab:Results} it is evident that the most of the errors are arising when there is a change in the system. For example, we can note that all the false alarms came for operation 6 and 7 are because of the change in the description of the operation. Moreover it can be noted that all the false positives are obtained only at these type of transitions only where a significant change is obtained in the description of the operation. We have also observed that the illegal sequence of operations are detected using the proposed method with a high recall rate of value more than 0.95 for most of the operations. But sometimes the machine is switched off abnormally at the end of $10^th$ operation and so most of the error came at this operation as the training is also not possible unless nose cleaning is done properly. On the other hand no label or ground truth is available. We have also observed that the overall wastage of resource due to miss use of of the machine can also be reduced by applying the proposed method. But as the Overall Equipment Effectiveness is not the focus of the current work we have not shown any result for that.

\begin{table*}[t]
  \caption{Sequence Wise Performance for the month of May}
  \label{tab:Results}
  \begin{tabular}{cccccccl}
    \toprule
    Operation &TP &FP &FN &TN &Recall &Precision &Accuracy (\%) \\
    \midrule
   
    1 &72	&0	&0	&3	&1	&1	&100\\
    2 &71   &0	&6	&3	&0.92	&1	&92.5\\
    3 &72   &0	&0	&1	&0.99	&1	&100\\
    4 &68   &0	&1	&3	&0.99	&1	&98.61\\
    5 &69   &0	&2	&2	&0.97	&1	&97.26\\
    6 &59   &4	&3	&8	&0.95	&0.94	&90.54\\
    7 &69   &3	&4	&5	&0.95	&0.96	&91.36\\
    8 &70   &2	&0	&1	&1	&0.97	&97.26\\
    9 &73   &1	&0	&3	&1	&0.99	&98.70\\
    10 &66   &3	&4	&1	&0.94	&0.96	&90.54\\
\\
  \bottomrule
\end{tabular}
\end{table*}

\section{Conclusions}
In this paper we have proposed a system that can raise an alarm in case there is a health hazard of the user may involve. It also helped the organization as a whole to save resource in terms of time by reducing the wrong usage of the machine. The proposed method used MDP and thus is adaptable to changes in the system by incorporating reward function. On the other hand it also used forgetting factor using an LRU algorithm to remove stale data and thus avoid the memory crunch of the system. In an IIoT scenario this adaptability is very much required so that it can automatically learn new concepts and changes without re-training it every time. Over and above the forgetting outdated concepts is also required to make a workable system. This proposed system works with an overall accuracy of 95.61\%, and Recall 0.97, Precision 0.98. The current work has a limitation that it uses the thresholds by applying some very simple statistics. In future this can be replaced by some sophisticated Machine Learning techniques. 

\section*{References}
%
%

[1] Gert Cauwenberghs and Tomaso Poggio. 2001. Incremental and decremental
support vector machine learning. In Advances in neural information processing
systems. 409–415.

[2] ISO, "Safety of machinery - Emergency stop function - Principles for design".
ISO.

[3] ISO, "safety of machinery - General principles for design - Risk assesment
and risk reduction". ISO.

[4] ISO, "Safety of machinery Positioning of the safeguards w.r.t the approach
speeds of parts of the human body". ISO.

[5] Imtiaz Ali Khan. Ergonomic Design of Human-CNC Machine Interface.
Human Machine Interaction Maurtua Inaki, IntechOpen. https://doi.org/10.5772/
26411

[6] Daniel Kifer, Shai Ben-David, and Johannes Gehrke. 2004. Detecting Change in
Data Streams. In Proceedings of the Thirtieth International Conference on Very Large
Data Bases - Volume 30 (VLDB ’04). 180–191.

[7] "TCS", "INTERNET OF THINGS". Retrieved Aug 17, 2018 from https:
//www.tcs.com/internet-of-things

[8] "Wikipedia", "Human machine interface". Retrieved Aug 17, 2018 from
$https://oshwiki.eu/wiki/Human_machine_interface$

\end{document}